\documentclass[11pt,a4paper]{article}
\usepackage{float}
\usepackage{epsfig}
\usepackage{multirow}
\usepackage[cp1252,utf8]{inputenc}
\usepackage{inputenc}
\usepackage{hyperref}
\usepackage{graphicx}
\usepackage{wrapfig}
\usepackage{amsmath}
\usepackage{xparse}
\usepackage{amsfonts}
\usepackage{amssymb}
\usepackage{relsize} 
\usepackage[top=1in, bottom=2cm, left=2cm, right=2cm]{geometry}
\usepackage{subcaption}
\usepackage{authblk}
\usepackage{ulem}
\usepackage{physics}
\usepackage{xcolor}
\NewDocumentCommand{\tens}{t_}
{%
	\IfBooleanTF{#1}
	{\tensop}
	{\otimes}%
}
\NewDocumentCommand{\tensop}{m}
{%
	\mathbin{\mathop{\otimes}\displaylimits_{#1}}%
}
\usepackage{changepage} 

\usepackage{afterpage}

\usepackage{placeins}

\usepackage{wrapfig}
\usepackage{cite} 
\usepackage[nottoc,notlof,notlot]{tocbibind} 

\title{\bf Gauss-Bonnet AdS planar and spherical black hole thermodynamics and holography}
\author{\bf  Souvik Paul \thanks{souvik.paul@bose.res.in}}
\author{\bf Sunandan Gangopadhyay \thanks{sunandan.gangopadhyay@bose.res.in}}
\author{\bf  Ashis Saha \thanks{ashis.saha@bose.res.in}}

\affil{Department of Astrophysics and High Energy Physics,\\
	S.N.~Bose National Centre for Basic Sciences,\\
	Salt Lake, Kolkata 700106, India}

\date{}

\begin{document}
	\maketitle
	\begin{abstract}
\noindent In this work, we extend the study in \cite{Bilic:2022psx} incorporating the AdS/CFT duality to establish a relationship between the local temperatures (Tolman temperatures) of a large (AdS) spherical and a (AdS) planar Schwarzschild black hole near the AdS boundary considering Gauss-Bonnet curvature correction in the gravitational action. We have shown that the higher curvature corrections appear in the local temperature relationship due to the inclusion of Gauss-Bonnet term in the bulk. By transforming the metric into Fefferman-Graham form, we have calculated the energy density of the conformal fluid at the boundary.  The obtained result contains finite coupling corrections which are holographically induced by the Gauss-Bonnet curvature correction in the bulk theory. Following the well known approach of fluid/gravity duality, the energy density of the conformal fluid at the boundary is then compared with the black body radiation energy density. This comparison shows that the energy density is proportional to the temperature of the conformal fluid. The temperature of the conformal fluid is then shown to be related to the Tolman temperature of the black hole which then eventually helps us to establish both the Hawking temperature and Tolman temperature relationship between large spherically symmetric and planar Schwarzschild black holes in Gauss-Bonnet gravity near the AdS boundary.
\end{abstract}

\section{Introduction}
\noindent The black hole solutions of the Gauss-Bonnet (GB) gravity have always been a matter of great interest due to their interesting topological properties \cite{PhysRevD.107.064023, PhysRevD.105.104053}. In the context of string theory, they emerge naturally due to the inclusion of the GB curvature correction\footnote{For any quantum theory of gravity, the presence of the curvature correction terms in the effective action are justifiedly expected.} \cite{PhysRevD.80.104032, PhysRevD.86.104016}. The thermodynamic properties of GB black holes, particularly in AdS spacetime \cite{PhysRevD.65.084014,PhysRevD.38.2434,Nojiri:2001aj,Nojiri:2002qn,Cvetic:2001bk,clunan2004gauss} has been getting lot of attention for the past few decades, due to their applicability in the context of the AdS/CFT duality \cite{Witten:1998qj,maldacena1999large,gubser1998gauge,witten2001anti,Natsuume:2014sfa,Nastase:2007kj}. These higher-dimensional entities exhibit new features, such as multiple horizons, which can profoundly affect their thermodynamic properties and stability. The GB parameter changes the definition of the volume along with the effects on the phase structure of the black hole especially near the critical point when mixed first law is considered. Furthermore, the GB coupling term also modifies the thermodynamic Smarr relation \cite{PhysRevD.107.046005,Karch:2015rpa,Cong:2021fnf,Cong:2021jgb}. Another interesting aspect of GB theory is that it  arises from the low-energy limit of heterotic string theory\cite{Zwiebach:1985uq,Gross:1986iv,Gross:1986mw,Metsaev:1987zx,Metsaev:1986yb}. It can be shown that the low energy effective action of heterotic string theory consists of GB term and one scalar field. Although the functional form of the scalar field remains a fascinating and challenging topic of research, some previous studies\cite{Metsaev:1986yb} have considered a trivial constant value of the scalar field, and therefore, one can ignore its effect on the solutions. In the context of the AdS/CFT duality, it has been noted that the Hawking-Page phase transition for the AdS black hole is in fact dual to the confinement-deconfinement transition for the strongly coupled dual gauge theory. In case of the AdS black hole solutions of the GB gravity, they undergo Hawking-Page transition to thermal GB AdS spacetime \cite{sahay2017geometry,dey2007phase,astefanesei2008attractor,cho2002anti,PhysRevD.76.024011,Hendi:2015pda,Hendi:2016njy,Hendi:2016yof}. They exhibit unique phase transition behaviour different from black hole solutions in Einstein gravity. The phase structure of these black holes has been extensively studied under different conditions, including extended phase space and with additional fields like dilaton, Maxwell or higher order gauge fields \cite{Hendi:2017fxp,el2018revisiting,liang2020phase,anninos2009thermodynamics,Hendi:2017lgb}. In the context of the AdS/CFT correspondence, one might question about the importance of the GB curvature correction in the bulk theory (gravity). In order to understand this, one must keep in mind the most basic essence of the AdS/CFT duality, which deals with the fact that an Einstein gravity dual exists for a SU$(\mathcal{N})$ gauge theory in the limit $\mathcal{N}\rightarrow\infty$ and $g_{\mathrm{YM}}\rightarrow 0$, where $\mathcal{N}$ represents the rank of the gauge group and $\lambda=g_{\mathrm{YM}}^2 \mathcal{N}\gg1$ (strong-coupling limit) is the 't Hooft coupling constant. On the other hand, it has been shown that the presence of higher curvature terms in the gravitational theory in fact represents $1/\mathcal{N}$ or $1/\lambda$ corrections for the dual gauge theory. This implies that one can denote the effects of including the higher curvature in the gravity theory as the finite 't Hooft coupling corrections for the gauge theory. For the sake of better understanding, we consider the famous holographic result of viscocity ($\eta$) to entropy density ($s$) ratio \cite{Policastro:2001yc,Kovtun:2003wp}
 \begin{equation}
\frac{\eta}{s} = \frac{1}{4\pi}~.
\label{kss ratio}
\end{equation}
The above result is sometimes denoted as the KSS (Kovtun-Son-Starinets) bound or the universal bound and it holds for the dual SU$(\mathcal{N})$ gauge theory in the limit $\mathcal{N}\rightarrow\infty$. On the other hand, it has been shown that if one includes the GB curvature corrections in the bulk theory, the result gets modified to the following form \cite{brigante2008viscosity}
\begin{equation}
	\frac{\eta}{s} = \frac{1}{4\pi} \left[1-2\frac{(d-1)}{(d-2)}\lambda\right]~.
\end{equation}
The above result indicates that for a positive value of the GB parameter $\lambda$, the KSS bound is getting violated. This interesting observation justifies to include higher curvature corrections to the bulk theory (or in the dual sense, finite coupling corrections for the boundary gauge theory) in the context of holographic investigations of various physical quantities. Some of the interesting studies in this direction can be found in \cite{bhattacharyya2008nonlinear,gangopadhyay2012analytic, GANGOPADHYAY2013176,bu2015hydrodynamics,hu2022ads,chandranathan2023entropy}. On the other hand, in context of the gauge/gravity duality, the topology of the bulk geometry (planar, spherical or hyperbolic) plays a crucial role as it depicts the symmetries of the field theory localized at the boundary. However, it is a well-known fact that there is a long standing issue with the thermodynamic variables of a planar black hole in AdS \cite{Horowitz:1998ha}. In order to have a grasp of this, we first consider a static, spherically symmetric Schwarzschild black hole in AdS$_{d+1}$ which can be characterized by the following metric \cite{Witten:1998qj,Horowitz:1998ha}
\begin{eqnarray}\label{eg1}
ds^2=-f_s(r)dt^2+\frac{dr^2}{f_s(r)}+r^2d\Omega^2_{d-1}~,~f_s(r)=\frac{r^2}{L^2}+1-\left(\frac{r_0M}{r^{d-2}}\right)
\end{eqnarray}
where the parameter $r_0=\frac{16\pi G_{d+1}}{(d-1)\Omega_{d-1}}$, $L$ is the AdS radius and $M$ is the mass of the black hole. For the sake of computational simplicity, one can also write down the above lapse function $f_s(r)$ as $f_s(r)=\frac{r^2}{L^2}+1-\mu_s\left(\frac{L}{r}\right)^{d-2}$, where $\mu_s=\frac{16\pi G_{d+1}}{(d-1)\Omega_{d-1}L^2}M$ \cite{Bilic:2022psx}. It can be easily shown that the Hawking temperature corresponding to the above black hole metric reads
\begin{eqnarray}\label{eg2}
T_H=	\frac{dr_+^2+(d-2)L^2}{4\pi r_+ L^2}
\end{eqnarray}
where $r_+$ is the radius of the event horizon which can be evaluated by solving the equation $f_s(r_+)=0$. Now for a large black hole ($r_+\gg L$), the above expression simplifies to
\begin{eqnarray}
	T_H=\frac{dr_+}{4\pi L^2}~.
\end{eqnarray}
The large black hole limit ($r_+\gg L$) is sometimes also denoted as the planar limit of a large spherically symmetric AdS black hole. It has also been suggested that for a large spherically symmetric AdS$_{d+1}$ Schwarzschild black hole, consideration of the planar limit changes the topology from $S^1\times S^{d-1}$ to $S^1\times R^{d-1}$ and keeping this in mind one can approximate the large black hole by a planar black hole \cite{Witten:1998qj}. As shown by Witten \cite{Witten:1998qj}, one can approximate the spherically symmetric metric (which corresponds to a large black hole) to the following planar metric by considering suitable scaling of the coordinates $t$ and $r$. This reads
\begin{eqnarray}
	ds^2=-\left(\frac{\rho^2}{L^2}-\frac{L^{d-2}}{\rho^{d-2}}\right)d\tau^2+\frac{d\rho^2}{\left(\frac{\rho^2}{L^2}-\frac{L^{d-2}}{\rho^{d-2}}\right)}+\rho^2\sum_{i=1}^{d-1}dx_i^2~.
\end{eqnarray}
The above metric represents a AdS$_{d+1}$ Schwarzschild black hole which has a planar horizon at $\rho=L$. Furthermore, the associated Hawking temperature is obtained to be
\begin{eqnarray}
	T_H=\frac{d}{4\pi L}~.
\end{eqnarray}
The above expression advocates for the fact that the Hawking temperature only depends upon the value of the AdS radius. However, since the value of the AdS radius $L$ is not unique, one cannot fix it which leads to an ill-defined Hawking temperature. This behaviour of the Hawking temperature is basically due to the translational invariance of the planar black hole horizon \cite{Horowitz:1998ha,hubeny2010hawking}. Similarly the temperature of extremal Reissner-Nordstrom black hole with AdS near horizon geometry is not well defined. In previous literatures \cite{Hawking:1995fd,Alvarenga:2003tx,Alvarenga:2003jd}, it was argued that the temperature of the extremal Reissner-Nordstr\"om black hole is arbitrary since its horizon is infinitely far away.
In a recent work \cite{Bilic:2022psx}, gauge/gravity duality has been used to resolve the horizon temperature ambiguity of a planar AdS-Schwarzschild (SAdS) black hole. However, as we have mentioned earlier, inclusion of higher curvature corrections in the bulk theory sometimes changes the observations that have been made for the pure Einstein gravity. Keeping this in mind, in this work we try to understand how the temperature of a planar SAdS black hole in GB gravity can be realized by using holography and what are effects of the finite coupling corrections on the result shown in \cite{Bilic:2022psx}. In particular we exploit the fact that the holographic stress tensor depends upon the near boundary expansion of the metric. Hence, we transform the bulk metric in Fefferman-Graham coordinates to obtain the near boundary expansion of the planar and spherical GB black hole. From this transformed metric, the appropriate component is read off. This is then needed to obtain the CFT stress tensor in the boundary. The result clearly reveals the dependence of the stress tensor on the GB parameter, and is one of the main findings in the paper. Using the expression for the stress tensor, we have showed the correspondence between the conformal fluid at the boundary and the geometries of the planar and spherical AdS GB black hole.\\
This paper is organized as follows. In section \eqref{Section2}, we have provided a brief discussion of the thermodynamic properties of AdS Gauss-Bonnet black holes with planar and spherical symmetry. In section \eqref{Section3}, we provide a short review on the gauge/gravity duality and discuss an expression for the holographic stress tensor of GB gravity.   In section \eqref{Section4}, thermodynamic quantities for a conformal fluid from gauge/gravity duality. The relationships between the local temperature of the planar and spherical black holes are also derived. Finally, we conclude in section \eqref{Section5}.

\section{Einstein-Hilbert action with Gauss-Bonnet correction}\label{Section2}
In this section, we look at the Einstein-Hilbert action with Gauss-Bonnet curvature correction in $(d+1)$ spacetime dimensions, with $d\geq4$. The action of this theory reads \cite{PhysRevD.65.084014, PhysRevD.107.046005,Deruelle:2017xel,hu2022ads} 
\begin{equation*}
	S =S_{bulk}+S_{boundary}
\end{equation*}
where
\begin{equation}
	S_{bulk}=\frac{1}{16 \pi G_{d+1}} \int_V d^{d+1}x \sqrt{\mid g \mid}\left[R-2 \Lambda +\frac{L^{2}\lambda}{(d-2)(d-3)}(R^2+R_{\mu \nu \rho \sigma}R^{\mu \nu \rho \sigma}-4 R_{\mu \nu}R^{\mu \nu})\right]
	\label{GB action}
\end{equation}
\begin{equation}\label{GHY}
	S_{boundary}=\frac{1}{8\pi G_{d+1}}\int_{\partial V} d^{d}y \sqrt{\mid h \mid}\left[K-T+\frac{2 L^{2} \lambda}{(d-2)(d-3)}(J-2 G^{ij}_{\partial V}K_{ij})\right]~.
\end{equation}
where $R$ is the Ricci scalar, $L$ is the curvature radius of $\text{AdS}_{d+1}$, $\Lambda=\frac{-d(d-1)}{L^{2}}$ is the cosmological constant, $\lambda$ is the GB coupling parameter and $\text{G}_{d+1}$ is the $(d+1)$ dimensional Newton's gravitational constant. The GB coupling parameter $\lambda$ is constrained within the range \cite{buchel2010holographic, buchel2009causality}
\begin{equation}
    \frac{-(d-2)(3d+2)}{4(d+2)^{2}} \leq \lambda \leq \frac{(d-2)(d-3)(d^{2}-d+6)}{4 (d^{2}-3d+6)^{2}}~.
\end{equation}
In the Gibbons-Hawking-York boundary term (given in eq.\eqref{GHY}), $h$ is the determinant of the induced metric, $G^{ij}_{\partial V}$ is the intrinsic Einstein tensor on the boundary $\partial V$, $T$ is the brane tension of the end of the world brane, $K_{ij}$ is the extrinsic curvature on the boundary $\partial V$ and $K=h^{ij}K_{ij}$ is its trace, $J$ is the trace of $J_{ij}$ given as
\begin{equation}
	J_{ij} = \frac{1}{3}\left(2 K_{ik} K^{k}_{j}-2 K_{ik} K^{kl} K_{lj} + K_{ij}(K_{kl}K^{kl}-K^{2})\right)~.
\end{equation} 
The GHY boundary term (\cite{PhysRevD.15.2752,PhysRevLett.101.031601,Deruelle:2017xel,Myers:1987yn}) in the above action gives a well defined variational problem. We shall see in the subsequent discussion that this term is needed to get the holographic stress-energy tensor.
 
\noindent
The gravitational action \eqref{GB action} leads to the following equation of motions
\begin{equation}
    R_{\mu \nu} - \frac{1}{2}R g_{\mu \nu} - \frac{d(d-1)}{2 L^{2}} g_{\mu \nu} + \frac{L^{2}\lambda}{(d-2)(d-3)} H_{\mu \nu} = 0
    \label{GB EOM}
\end{equation}
where $H_{\mu \nu}$ is given by
\begin{equation}
    H_{\mu \nu} = 2\left(R_{\mu \rho \sigma \zeta} R_{\nu}^{ \rho \sigma \zeta} - 2 R_{\mu \rho \nu \sigma}R^{\rho \sigma} + R R_{\mu \nu}\right) - \frac{1}{2}\left(R_{\alpha\beta\rho\sigma}R^{\alpha\beta\rho\sigma}- 4 R_{\alpha\beta}R^{\alpha\beta} + R^{2}\right)g_{\mu \nu}~.
\end{equation}

Now to solve the above equation we will choose the following ansatz for the metric
\begin{equation}
	ds^2 = -e^{2\nu}dt^{2} + e^{2\lambda}dr^{2} + r^2 h_{ij} dx^{i} dx^{j}\label{ansatz}
\end{equation}
where $\nu$ and $\lambda$ are functions of $r$ only, and $h_{ij} dx^{i} dx^{j}$; $i,j=1,2,...,d-1,$ represents the line element of a 
$(d-1)$-dimensional hypersurface having constant curvature $(d-1)(d-2)k$ and volume $\omega_{k}$\cite{Cai:2009ua}. It is to be mentioned that $k$ can take the values $1$, $0$ and $-1$ which represents spherical, planar and hyperbolic geometries respectively. The expression for $\omega_{k}$ reads\footnote{For $k=0$, the explicit expression of the volume of a $(d-1)$-dimensional hypersurface with constant $r$ and linear size $L_{1}$ is given by 	$\omega_{0}=\left(\int_{-L_{1}/2}^{L_{1}/2}dx\right)^{d-1} =L^{d-1}_{1}$\cite{witten2001anti}.}
\begin{equation}
	\omega_{k}=
	\begin{cases}
		L^{d-1}_{1},&k=0\\
		\frac{2\pi ^{d/2}}{\Gamma(d/2)},&k=1~.
	\end{cases}
\end{equation}

\noindent
The explicit form of this $(d-1)$-dimensional hypersurface line element reads\cite{Huang:2017ohr}
\begin{equation}
	h_{ij} dx^{i} dx^{j}=
	\begin{cases}
		d\theta^{2}_{1}+d\theta^{2}_{2}+\dots+d\theta^{2}_{d-1},&k=0\\
		d\theta^{2}_{1}+sin^{2}\theta_{1}(d\theta^{2}_{2}+sin^{2}\theta_{2}(d\theta^{2}_{3}+\dots+sin^{2}\theta_{d-2}d\theta^{2}_{d-1})),&k=1\\
		d\theta^{2}_{1}+sinh^{2}\theta_{1}(d\theta^{2}_{2}+sin^{2}\theta_{2}(d\theta^{2}_{3}+\dots+sin^{2}\theta_{d-2}d\theta^{2}_{d-1})),&k=-1~.
	\end{cases}
\end{equation}

\noindent
 Substituting eq.\eqref{ansatz} in eq.\eqref{GB EOM} leads to the following form of the lapse function \cite{PhysRevD.65.084014}
\begin{equation}
	e^{2\nu}=e^{-2\lambda}=f(r)= k+\frac{r^2}{2\lambda L^2}\left(1 \mp \sqrt{1-\frac{64\pi G_{d+1}L^{2}\lambda M}{(d-1)\omega_{k}r^{d}}-4\lambda}\right)\label{e mu nu}~
\end{equation}
where $M$ is the Arnowitt-Deser-Misner (ADM) mass\cite{PhysRev.116.1322}. From the above equation, one can observe that there are two possible solutions, one with positive sign and another with negative sign. By considering the propagation of graviton, in \cite{boulware1985string} it was shown that the solution with positive sign is unstable. On the other hand, the solution with negative sign is stable and the theory is ghost free. 


\noindent
The lapse function $f(r)$ in eq.\eqref{e mu nu} can be written in the following form 
\begin{equation}
    f(r) = k+\frac{r^2}{2\lambda L^2}\left(1 - \sqrt{1-4\lambda \left(1-\frac{m}{r^d}\right)}\right)~.
    \label{k lapse function}
\end{equation}
In the above equation the mass parameter $m$ is related with the Arnowitt-Deser-Misner (ADM) mass as 
\begin{equation}\label{ADM mass}
    M = \frac{(d-1) \omega_{k}}{16 \pi L^{2} G_{d+1}} m~.
\end{equation}

In the asymptotic limit ($r\to \infty$) (that is the near boundary limit), one obtains
\begin{equation}
    f(r) \approx\frac{r^2}{2\lambda L^2}\left(1 - \sqrt{1-4\lambda} \right)~.
\end{equation}
This observation naturally sets an effective radius for the $AdS$ spacetime as $L_{eff}^{2} = \frac{2\lambda L^{2}}{1 - \sqrt{1-4\lambda}}$ for $\lambda  \leq \frac{1}{4}$. The value $\lambda = \frac{1}{4}$ is known as the Chern-Simons limit \cite{gangopadhyay2012analytic}. As we intend to carry out the calculations in the near boundary limit, we stick to $\lambda \leq \frac{1}{4}$ throughout the work as in this range $f_\infty \equiv \lim_{r\to\infty} f(r) = \frac{1-\sqrt{1-4\lambda}}{2\lambda}$ is real.

\noindent
The mass parameter of the black hole can be expressed in terms of the horizon radius $r_{+}$. This can be seen as follows by setting $f(r_{+})=0 $ in eq.\eqref{k lapse function} gives 
\begin{equation}
	k+\frac{r_{+}^{2}}{2 \lambda L^{2}}\left(1-\sqrt{1-4\lambda(1-\frac{m}{r_{+}^{d}})}\right) = 0~.
	\label{f(r+)}
\end{equation}
Solving this for $m$, we get
\begin{eqnarray}
	1-4 \lambda (1-\frac{m}{r_{+}^{d}})=(1+\frac{2 \lambda L^{2} k}{r_{+}^{2}})^{2}\nonumber\\
	\implies\frac{4 \lambda m}{r_{+}^{d}}=4 \lambda +\frac{4 \lambda L^{2}k}{r_{+}^{2}}+\frac{4 \lambda^{2} L^{4}k^{2}}{r_{+}^{4}}\nonumber\\
	\implies m=r_{+}^{d-4} \left(r_{+}^{4} + L^{2} k r_{+}^{2} + \lambda L^{4} k^{2}\right)~.
	\label{mass general}
\end{eqnarray}
By using the above relation in eq.\eqref{ADM mass}, one can obtain the ADM mass in terms of the horizon radius. This reads
\begin{equation}\label{ADM_horizon}
	 M = \frac{(d-1) \omega_{k}}{16 \pi L^{2} G_{d+1}} r_{+}^{d-4} \left(r_{+}^{4} + L^{2} k r_{+}^{2} + \lambda L^{4} k^{2}\right)~.
\end{equation}
On the other hand, the Hawking temperature of the black hole reads\cite{Hawking:1982dh}
\begin{align}
	T &= \frac{1}{4 \pi }\frac{\partial f}{\partial r} \Bigg|_{r = r_{+}}\nonumber\\
    & = \frac{dr_{+}}{4\pi L^{2} (1+\frac{2kL^{2}\lambda}{r_{+}^{2}})} + \frac{k(d-2)}{4\pi r_{+} (1+\frac{2kL^{2}\lambda}{r_{+}^{2}})} + \frac{\lambda k^{2} L^{2}(d-4)}{4\pi r_{+}^{3} (1+\frac{2kL^{2}\lambda}{r_{+}^{2}})}~.
    \label{Temparature general}
\end{align}
The Bekenstein-Hawking entropy of the black hole must obey the first law of thermodynamics $dM = T dS$ \cite{Hawking:1982dh}. This in fact helps us to calculate the entropy of the black hole by using the notion of Hawking temperature and ADM mass. This turns out to be
\begin{align}
    S &=\int \frac{dM}{T}\nonumber \\&= \int_{0}^{r_{+}} \frac{1}{T} \left(\frac{\partial M}{\partial r_{+}}\right) dr_{+}\nonumber\\
    &=\frac{\omega_{k}r_{+}^{d-1}}{4 G_{d+1}}\left(1 + \frac{2 k L^{2}\lambda}{r_{+}^{2}}\frac{(d-1)}{d-3}\right)~
    \label{entropy general}
\end{align}
where in getting the third line of the equality we have used
\begin{equation}
    \frac{\partial M}{\partial r_{+}} = \frac{(d-1) \omega_{k} r_{+}^{d-5}}{16 \pi L^{2}  G_{{d+1}}}\left[d r_{+}^{4} + k L^{2}(d-2)r_{+}^{2} + \lambda k^{2} L^{4}(d-4)\right]~.
\end{equation}
In the upcoming sections we will denote $\omega_{k}$ as $\omega_{pl}$ for $k=0$ and $\omega_{k}$ as $\omega_{sp}$  for $k=1$.

\subsection{Planar AdS Gauss-Bonnet black hole}
As we are interested in the planar black hole solution, we now consider the scenario for which $k=0$. In this case, eq.(s) (\eqref{ADM_horizon}, \eqref{Temparature general},\eqref{entropy general}) boil down to the following forms respectively
\begin{equation}
	M_{pl} = \frac{(d-1) \omega_{pl}}{16 \pi L^{2} G_{d+1}} r_{pl}^{d}~,
	\label{M Planar}
\end{equation}
\begin{equation}
    T_{pl} = \frac{d r_{pl}}{4 \pi L^{2}}~,
    \label{T Planar}
\end{equation}
\begin{equation}
    S_{pl} = \frac{\omega_{pl}}{4 G_{d+1}} r_{pl}^{d-1}~
    \label{S Planar}
\end{equation}
where $\omega_{pl}$ is the volume of $(d-1)$-dimensional hypersurface with zero curvature. 
In the above expressions, $M_{pl}$, $T_{pl}$, and $S_{pl}$ are ADM mass, temperature and entropy of a planar SAdS black hole in Gauss-Bonnet gravity. Here $r_{pl}$ is the horizon radius of the planar black hole which is related to the black hole mass parameter as $r_{pl}=m^{\frac{1}{d}}$, which can be observed from the relation given in eq.\eqref{mass general}.

\noindent
The term $\omega_{pl}~r_{pl}^{d-1}$ is equivalent to the area of the event horizon of the black hole in $(d+1)$-dimensional AdS spacetime.
Hence we can say that the entropy of the planar black hole solution satisfies the area law \cite{bekenstein1973black}. On the other hand,  the expression of entropy density ($s_{pl}$) is given by\cite{buchel2010holographic}  
\begin{equation}
	s_{pl} = \frac{\omega_{pl}}{4 G_{d+1}} \left(\frac{r_{pl}}{L}\right)^{d-1}~.
\end{equation}
The above expression at $r_{pl}=L$ reduces to $s_{pl} = \frac{\omega_{pl}}{4 G_{d+1}}$\cite{Bilic:2022psx}.

\subsection{Spherical AdS Gauss-Bonnet black hole}
For $k = 1 $, the black hole solution has spherical symmetry which is characterized by the following thermodynamic quantities
\begin{equation}
    M_{sp} = \frac{(d-1) \omega_{sp}}{16 \pi L^{2} G_{d+1}}r_{sp}^{d} \left(1 + \frac{L^{2}}{r_{sp}^{2}}  + \frac{\lambda L^{4}}{r_{sp}^{4}} \right)~
    \label{M spherical}
\end{equation}
\begin{equation}
	T_{sp} = \frac{dr_{sp}}{4\pi L^{2} (1+\frac{2L^{2}\lambda}{r_{sp}^{2}})} + \frac{(d-2)}{4\pi r_{sp} (1+\frac{2L^{2}\lambda}{r_{sp}^{2}})} + \frac{\lambda ^{2} L^{2}(d-4)}{4\pi r_{sp}^{3} (1+\frac{2L^{2}\lambda}{r_{sp}^{2}})}~
	\label{T spherical}
\end{equation}
\begin{equation}
    S_{sp} = \frac{\omega_{sp}}{4 G_{d+1}} r_{sp}^{d-1}\left(1 + \frac{2  L^{2}\lambda}{r_{sp}^{2}}\frac{(d-1)}{d-3}\right)~
    \label{S spherical}
\end{equation}
where $\omega_{sp}$ is the volume of $(d-1)$ dimensional hypersurface with constant curvature $(d-1)(d-2)$, and the expression of $\omega_{sp}$ reads
\begin{equation}
	\omega_{sp}=\frac{2\pi ^{d/2}}{\Gamma(d/2)}~.
\end{equation}
\noindent
For large black holes, one can consider $r_{sp}\gg L$. Under this condition, the quantities assume the following simpler forms
\begin{equation}
    T_{sp} \approx \frac{d r_{sp}}{4 \pi L^{2} }~
    \label{Approx T Spherical }
\end{equation}
\begin{equation}
    M_{sp} \approx \frac{(d-1) \omega_{sp}}{16 \pi L^{2} G_{d+1}}r_{sp}^{d}~
    \label{Approx M Spherical }
\end{equation}
\begin{equation}
    S_{sp} \approx \frac{ \omega_{sp}}{4 G_{d+1}} r_{sp}^{d-1}~.
    \label{Approx S Spherical }
\end{equation}

\noindent
In the above expressions $M_{sp}$, $T_{sp}$ and $S_{sp}$ are the ADM mass, temperature and entropy of a spherical SAdS black hole in GB gravity. Here $r_{sp}$ is the horizon radius which is related to the black hole mass parameter as 
\begin{equation}
	 m=r_{sp}^{d-4} \left(r_{sp}^{4} + L^{2} r_{sp}^{2} + \lambda L^{4} \right)\label{mass parm spherical}~.
\end{equation}
\begin{figure}[H]
	\centering
	\includegraphics[width=11cm]{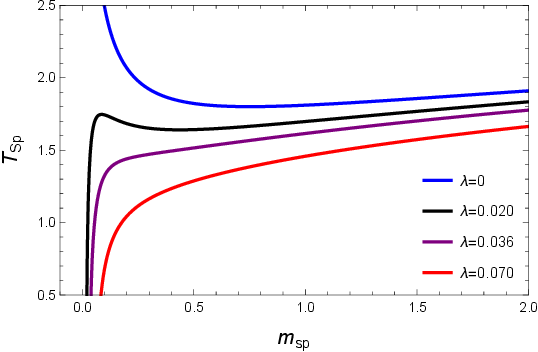}
	\caption{$T_{sp}$ vs $m_{sp}$ plot for different values of the GB coupling coefficient ($\lambda$) and $d=4$, $L=1$.}
	\label{Tvsm}
\end{figure}

\noindent
In Fig.\eqref{Tvsm}, we have plotted the variation of spherical GB black hole temperature ($T_{sp}$) with it's mass ($m_{sp}$). This plot has been done for $d=4$ and $L=1$. From the plot we can see that the Hawking temperature has a peak that shifts to the right for increasing value of $\lambda$. For smaller and smaller values of $\lambda$ the temperature peak incre when the horizon radius becomes zero. This observation is indeed consistent as in the limit $\lambda\to 0$ the GB gravity becomes Einstein gravity and the associated black hole solution is Schwarzschild black hole, which has diverging temperature near the point with vanishing horizon radius. It is also clear from this plot that the divergence of black hole temperature at the origin is regulated by GB coupling parameter.
\section{Gauge/gravity correspondence}\label{Section3}
The gauge/gravity duality in its first avatar was shown by Juan Maldacena in his seminal paper\cite{maldacena1999large}. It was shown that a type IIB superstring theory on AdS$_5\times S^5$ is equivalent to a $3+1$-dimensional $\mathcal{N}=4$ super Yang-Mills (SYM) theory in the large $\mathcal{N}$ limit \cite{maldacena1999large}. Furthermore, it was also shown that the  $\mathcal{N}=4$ SYM theory is a strongly coupled ($\lambda\gg1$) and conformally invariant theory. This striking observation later on helped us to summarize the fact that gauge/gravity duality relates the bulk geometry of a weakly interacting gravity theory to a strongly interacting gauge theory living at the boundary.\\
To study the gauge theory at the AdS boundary, we need to transform both the planar and spherical AdS Gauss-Bonnet metric in Fefferman-Graham coordinates. In Fefferman-Graham coordinates any asymptotic AdS metric takes the form \cite{zbMATH03971640,zhang2015holographic}
\begin{equation}\label{FG}
	ds^2 = \frac{L^{2}}{z^2}\left[g_{ab}(z,\Bar{x})dx^{a} dx^{b} + dz^2 \right]
\end{equation}
where the $d$-dimensional metric $g_{ab}$ can be expanded near the boundary ($z=0$) as  
\begin{equation}
	g_{ab} = g_{ab}^{(0)} + g_{ab}^{(2)} z^{2} + \dots + g_{ab}^{(d)} z^{d} + h_{ab}^{(d)} z^{d} \ln{z^{2}} + O(z^{d+1})~,~a,b=0,1,2 \dots (d-1)
	\label{metric power series}
\end{equation}

\noindent
It is important to note that any asymptotically AdS metric can be brought to the form given above near the boundary.

\noindent
The Fefferman-Graham form of the metric allows the vacuum expectation value of the stress tensor to be written as\cite{deHaro:2000vlm,Balasubramanian:1999re}
 \begin{equation}
    \left<T_{ab}\right> =  \frac{d L^{d-1}}{16 \pi G_{N}} g_{ab}^{(d)} + X_{ab}[g^{(n)}]~
    \label{stress tensor Einstein}  
 \end{equation}

\noindent
where $X_{ab}[g^{(n)}]$ is a function of $g^{(n)}$ with $n<d$. Its exact form depends on the spacetime dimension and it reflects the conformal anomalies of the boundary CFT\cite{Henningson:1998gx}. When $d$ is odd there are no gravitational conformal anomalies hence $X_{ab}=0$ \cite{deHaro:2000vlm} \cite{Gangopadhyay:2020xox}. We therefore note that to calculate the dual stress tensor, it is sufficient to know only the terms $g^{(0)}$ and $g^{(n)}$ with $n<d$. The coefficient of the logarithmic term, $h_{ab}^{(d)}$ in \eqref{metric power series} is directly proportional to the metric variation of conformal anomaly \cite{Bilic:2022psx}. The first term in eq.\eqref{stress tensor Einstein} gives the area law contribution to the black hole entropy and the second term($X_{ab}[g^{(n)}]$) correspond to the logarithmic correction to the black hole entropy\cite{Cai:2009ua,Chang:2018pnb,Fursaev:1994te}, when $d$ is even. Since we are interested only in the area law contribution hence we neglect the second term in our analysis. 

\noindent
In recent times, another method has been proposed to calculate the holographic stress tensor arising from higher derivative gravity dual using the first law of entanglement \cite{faulkner2014gravitation,bhattacharya2013thermodynamical}. For higher curvature gravity, the expression for the holographic stress tensor vacuum expectation value (VEV) reads \cite{sen2014holographic}
 \begin{equation}
     \left<T_{ab}\right> = d L^{d-3}\left[c_{1} + 2(d-2) c_{6}\right] g_{ab}^{(d)}~
 \end{equation}
 where $c_{1}$ is proportional to the $A$-type Euler anomaly and $(c_{1} + 2(d-2) c_{6})$ is proportional to the $B$-type Euler anomaly\cite{sen2014holographic}.
 
 \noindent
  In case of Gauss-Bonnet gravity, the VEV of the holographic stress-energy tensor has been obtained in \cite{hu2022ads}. Here we shall start by providing a brief discussion on the derivation of this result.
\noindent  
At first we will take the variation of the total action in eq.\eqref{GB action} and mainly focus on the boundary term, which gives
  \begin{equation}
  	\delta S = \frac{1}{2} \int_{\partial V} d^{d}y \sqrt{\mid h \mid} T^{ij}_{GB} \delta h_{ij}~
  \end{equation}
 where $T^{ij}_{GB}$ is the Brown-York stress tensor\cite{PhysRevD.47.1407} for Gauss-Bonnet gravity, given by
 \begin{equation}
 	T^{ij}_{GB}=-\frac{1}{8 \pi G_{d+1}}\left[K^{ij}-(K- T)h^{ij}+\frac{2L^2 \lambda}{(d-2)(d-3)}(Q^{ij}-\frac{1}{3}Q h^{ij})\right]~.
 \end{equation}
 Here $Q$ is the trace of $Q_{ij}$ where $Q_{ij}$ is given by
 \begin{equation}
 	Q_{ij}=3J_{ij} + 2K R_{Qij}+R_{Q}K^{ij}-2K^{kl}R_{Qkijl}-4R_{Qk(i}K_{j)}^{k}~.
 \end{equation}
 For a well defined variational problem one can impose various boundary conditions \cite{Witten:2018lgb,York:1972sj,Papadimitriou:2005ii,Anderson:2010ph,Anderson:2006lqb}. Application of Neumann boundary condition leads to \cite{hu2022ads}
 \begin{equation}
 	K^{ij}-(K- T)h^{ij}+\frac{2L^2 \lambda}{(d-2)(d-3)}(Q^{ij}-\frac{1}{3}Q h^{ij})=0~.
 \end{equation}
Now rewriting the action by applying the background field method is useful to study Weyl anomaly and correlation functions.
Hence we shall expand the action (eq.\eqref{GB action}) in terms of the background curvature $\bar{R}$, which is defined as
\begin{eqnarray}
	R = \bar{R}-\frac{d(d+1)}{l^2}\label{background curvature1}~~~~~~~~~~~~~\\
	R_{\mu \nu}=\bar{R}_{\mu \nu}-\frac{d}{l^2} g_{\mu \nu}\label{background curvature2}~~~~~~~~~~~~\\
	R_{\mu \nu \rho \sigma}=\bar{R}_{\mu \nu \rho \sigma}-\frac{1}{l^2} (g_{\mu \rho} g_{\nu \sigma} - g_{\mu \sigma} g_{\nu \rho})\label{background curvature3}~.
\end{eqnarray} 
Using the above relations, we get the expression of the action (eq.\eqref{GB action}) to be
\begin{equation}
	\begin{split}
		S&=\frac{1}{16 \pi \bar{G}_{d+1}} \int_V d^{d+1}x \sqrt{\mid g \mid}\left[R-2 \Lambda +\alpha l^{2} \mathcal{L}_{GB}(\bar{R})\right]\\&+\frac{1}{8\pi \bar{G}_{d+1}}\int_{\partial V} d^{d}y \sqrt{\mid h \mid}\left[(1+ 2\alpha (d-1)(d-2))(K-T)+2 \alpha l^{2}(J-2 G^{ij}_{\partial V}K_{ij})\right]~
	\end{split}
	\label{GB action reparameter}
\end{equation}
where $\mathcal{L}_{GB}(\bar{R})$ is the Gauss-Bonnet term (given in eq.\eqref{GB action}) which is represented in terms of the background curvature in eq(s).(\eqref{background curvature1},\eqref{background curvature2},\eqref{background curvature3}). 
In order to obtain the above eq.\eqref{GB action reparameter}, we have re-parameterized the AdS radius ($L$), Newton's constant ($G_{d+1}$) and the Gauss-Bonnet parameter ($\lambda$) in the following way
\begin{align}
	&\frac{1}{L^{2}}=\frac{1+\alpha (d-2)(d+1)}{1+ 2\alpha (d-1)(d-2)}\frac{1}{l^{2}}~,\label{L reparameter}\\
	& \frac{1}{G_{d+1}} = \frac{1+2\alpha (d-1)(d-2)}{\bar{G}_{d+1}}~,\label{G reparameter}\\
	&\frac{\lambda}{(d-2)(d-3)}=\frac{1+\alpha (d-2)(d+1)}{(1+2 \alpha (d-1)(d-2))^{2}} \alpha~.
	\label{alpha reparameter}
\end{align}
In eq.(s)(\eqref{L reparameter},\eqref{G reparameter},\eqref{alpha reparameter}), the
 parameter $\alpha$ is constrained within the range \cite{hu2022ads}
\begin{equation}
	\frac{1}{8}\geq\alpha \geq \frac{-1}{4(d^2-2d -2)}\label{alpha bound}~.
\end{equation} 
Now the Neumann boundary condition reads
\begin{equation}
	(1+2 \alpha (d-1)(d-2))(K^{ij}-(K-T)h^{ij})+ 2 \alpha l^{2}(Q^{ij}-\frac{1}{3}Q h^{ij})=0~.
\end{equation}
Variation of the above action (eq.\eqref{GB action reparameter}) leads to the same equation of motion (eq.\eqref{GB EOM}), although the free energy has contributions from the GHY boundary term, hence the vacuum expectation value of the holographic stress-energy tensor has corrections from the GHY boundary term. Taking the tracelessness in account, the vacuum expectation value of a holographic stress tensor with Gauss-Bonnet curvature correction can be calculated as \footnote{A detailed calculation of the holographic stress tensor for Gauss-Bonnet gravity is given in \cite{sen2014holographic}. } 
\begin{equation}
     \left<T_{ab}\right> = d \left[1 + 4 (d-2) \alpha\right] g_{ab}^{(d)}~.
     \label{gauusbonnet stress tensor}
 \end{equation}
where $g_{ab}^{(d)}$ is defined in eq.\eqref{metric power series}. The above equation is crucial in establishing the connection between the VEV of the stress tensor of the CFT and the bulk geometry. For simplicity we have assumed $16\pi \bar{G}$ and $l$ equals to one.

\noindent
In the upcoming subsections, we will use eq.\eqref{gauusbonnet stress tensor} to study the correspondence between the VEV of CFT stress tensor and the geometries of the planar and large spherical AdS Gauss-Bonnet black hole.
\subsection{Planar AdS Gauss-Bonnet black hole}
In this subsection, we will now discuss the correspondence between the VEV of the CFT stress tensor and the geometry of the planar SAdS black hole of GB gravity. In order to do this, the first step is to expand the lapse function (given in eq.\eqref{k lapse function}) upto $\mathcal{O}(\lambda)$ for $k=0$. This gives
\begin{equation}
	f(r) = \frac{r^2}{L^2}\left[\left(1-\frac{r_{pl}^d}{r^d}\right) +  \lambda\left(1-2 \frac{r_{pl}^d}{r^d}\right)\right]~.
\end{equation}
We now choose a suitable coordinate transformation $r = \frac{\Tilde{L}^2}{\rho}$ in order to transform the metric to the following form
\begin{equation}
    ds^2 = \frac{\Tilde{L}^{2}}{\rho^2}\left[-F(\rho) dt^2 + \frac{1}{F(\rho)} d\rho^2 + \Tilde{L}^2\sum_{i=1}^{d-1} (dx^{i})^{2}\right]
    \label{metric in inverse cordinate}
\end{equation}
where upto $\mathcal{O}(\lambda)$, $F(\rho)$ is given by
\begin{equation}
	F(\rho)=\left(1- (1+\lambda)\frac{\rho^{d}r_{pl}^{d}}{\Tilde{L}^{2d}}\right)~,
	\label{F(rho)}
\end{equation}
 $dx^{2}=\sum_{i=1}^{d-1} (dx^{i})^{2}$ and $\Tilde{L}$ is the modified $AdS$ radius in eq.\eqref{metric in inverse cordinate} such that it is related to the $AdS$ radius $L$ as   $\Tilde{L}^2 =\frac{L^2}{(1+\lambda)}$.

\noindent
We now proceed to transform the above line element to the Fefferman-Graham form given in eq.\eqref{FG}. This is the crucial point in our analysis as this will allow us to obtain the form of $g^{(d)}_{ab}$. This result is required in obtaining the CFT stress tensor (using eq.\eqref{gauusbonnet stress tensor}). In order to do this, we apply the following coordinate transformation 
\begin{align}
        \frac{dz}{z} &= \frac{d\rho}{\rho \sqrt{F(\rho)}}~\nonumber\\& =\frac{d\rho}{\rho \sqrt{1- (1+\lambda)\frac{\rho^{d}r_{pl}^{d}}{\Tilde{L}^{2d}}}}~,\nonumber\\
        &= \left[\frac{d\rho}{\rho}+(1+\lambda)\frac{\rho^{d-1} r_{pl}^{d}}{2 \Tilde{L}^{2d}}d\rho\right]+O(\lambda^{2})~. 
        \label{dz/z planar}
\end{align}
Integrating the above equation near the $AdS$ boundary ($\rho \to 0 $) and considering the expression upto $O(\lambda)$, we get
\begin{equation}
	\ln z=\ln \rho + (1+\lambda)\frac{\rho^{d} r^{d}_{pl}}{2d \tilde{L}^{2d}}~.
\end{equation}
Near the $AdS$ boundary ($\rho \to 0 $), the above equation can be recast in the form (upto $O(\lambda)$)  
\begin{equation}
	\rho=z\left[1-(1+\lambda)\frac{z^{d} r^{d}_{pl}}{2d \tilde{L}^{2d}}\right]~.
	\label{rho in z planar}
\end{equation}
Substituting this in eq.(\eqref{metric in inverse cordinate},\eqref{F(rho)}) gives
\begin{equation}    
        ds^{2} = \frac{\Tilde{L}^{2}}{z^{2}}\left[(-dt^{2} + \Tilde{L}^{2} \sum_{i=1}^{d-1}(dx^{i})^{2}) z^{0} + \frac{(1+\lambda) r_{pl}^{d}}{d \Tilde{L}^{2d}} ((d-1)dt^{2} +\Tilde{L}^{2}\sum_{i=1}^{d-1}(dx^{i})^{2})z^{d} + dz^{2}\right]~.
        \label{planar fefferman graham form}
\end{equation}
It can be seen that the above form of the line element has a term that is independent of $z$ followed by $z$ dependent terms which become zero in the limit $z \to 0$. Thus, the above form of the line element is identical to the standard Fefferman-Graham form (given in eq.(s)(\eqref{FG},\eqref{metric power series}). By comparing eq.\eqref{planar fefferman graham form} and eq.\eqref{FG}, one can determine the coefficients of $z^0$ and $z^{d}$, and this read
\begin{equation}
    g_{ab}^{(0)} dx^{a} dx^{b} = -dt^{2} + \Tilde{L}^{2} \sum_{i=1}^{d-1} (dx^{i})^{2}~
\end{equation}
\begin{equation}
	g_{ab}^{(d)} dx^{a} dx^{b} = \frac{(1+\lambda) r_{pl}^{d}}{d \Tilde{L}^{2d}} \left((d-1)dt^{2} +\Tilde{L}^{2}\sum_{i=1}^{d-1} (dx^{i})^{2}\right)~.
	\label{gabd planar}
\end{equation}
We will use the result in eq.\eqref{gabd planar} in our subsequent discussion to obtain the CFT stress tensor. 
\subsection{Spherical AdS Gauss-Bonnet black hole}
In this section, we derive the Fefferman-Graham form of the line element for the spherical AdS Gauss-Bonnet black hole.
Setting $k=1$ in eq.\eqref{k lapse function}, we note that the lapse function of a SAdS black hole in GB gravity with spherical symmetry is given by 
\begin{equation}
     f(r) = 1 + \frac{r^2}{2\lambda L^2}\left(1-\sqrt{1-4\lambda \left(1-\frac{m}{r^d}\right)}\right)~.
     \label{spherical lapse function}
\end{equation}
Similar to the planar case, we now expand the above lapse function upto $\mathcal{O}(\lambda)$. This gives
\begin{equation}
    f(r)= 1 + \frac{r^2}{L^2}\left[\left(1-\frac{r_+^d}{r^d}\right) +  \lambda\left(1-2 \frac{r_+^d}{r^d}\right)\right]~.
    \label{lapse lambda planar}
\end{equation}
Furthermore, in terms of the modified $AdS$ radius $\Tilde{L}$ the lapse function becomes
\begin{equation}
    f(r)\approx \frac{r^{2}}{\Tilde{L}^{2}}\left[1 + \frac{\Tilde{L}^{2}}{r^{2}} - \frac{m}{r^{d}} (1+\lambda)\right]~.
    \label{expanded spherical lapse function}
\end{equation}
By following the same procedure we have shown for the planar case, once again we introduce the coordinate transformation $r=\frac{\Tilde{L}^{2}}{\rho}$ and obtain the following form of the metric
\begin{equation}
    ds^2 = \frac{\Tilde{L}^{2}}{\rho^2}\left[-F(\rho) dt^2 + \frac{1}{F(\rho)} d\rho^2 + \Tilde{L}^{2}d\Omega_{d-1}^{2}\right]~.
    \label{spherical inverse coordinate}
\end{equation}
where $F(\rho)$ upto $\mathcal{O}(\lambda)$ has the form
\begin{equation}
	F(\rho)  = \left(1 + \frac{\rho^{2}}{\Tilde{L}^{2}} -(1+\lambda)\frac{m \rho^{d}}{\Tilde{L}^{2d}}\right)~.
\end{equation}
Once again we proceed to obtain the Fefferman-Graham form for the above metric. This we do by using the following coordinate transformation
\begin{align}
        \frac{dz}{z} &= \frac{d\rho}{\rho \sqrt{F(\rho)}}\nonumber \\&=\frac{d\rho}{\rho \sqrt{1 + \frac{\rho^{2}}{\Tilde{L}^{2}} -(1+\lambda)\frac{m \rho^{d}}{\Tilde{L}^{2d}}}}~,\nonumber\\
        &= \frac{d\rho}{\rho} - \frac{\rho^{2}}{2 \Tilde{L}^{2}} d\rho + \frac{3 \rho^{3}}{8 \Tilde{L}^{4}} d\rho + m(1+\lambda) \frac{\rho^{d-1}}{2\Tilde{L}^{2d}}d\rho +O(\lambda^{2})~.
        \label{dz/z spherical}
\end{align}
Integrating the above equation near the $AdS$ boundary ($\rho \to 0 $) and considering the expression upto $O(\lambda)$, we get
\begin{equation}
	\ln z=\ln \rho -\frac{\rho^{2}}{4\tilde{L}^{2}}+\frac{3}{32}\frac{\rho^{4}}{\tilde{L}^{4}}+m(1+\lambda)\frac{\rho^{d}}{2d\tilde{L}^{2d}}~.
	\label{lnz lnrho}
\end{equation}
Finally the above equation can be recast in the form
\begin{equation}
	\rho=z\left[1+\frac{z^{2}}{4\tilde{L}^2}-\frac{z^{4}}{16\tilde{L}^4}-(1+\lambda)\frac{m z^{d}}{2d\tilde{L}^{2d}}\right]~.
	\label{z in rho spherical}
\end{equation}
Substituting this in eq.\eqref{spherical inverse coordinate}, one obtains the line element(eq.\eqref{spherical inverse coordinate}) in the desired Fefferman-Graham form as
\begin{equation}
    ds^{2} = \frac{\Tilde{L}^{2}}{z^{2}}\left[-A(z)dt^{2} + B(z) \Tilde{L}^{2} d\Omega_{d-1}^{2} +dz^{2}\right]
    \label{fefferman spherical }
\end{equation}
where the expressions for $A(z)$ and $B(z)$ read
\begin{eqnarray}
    A(z) &=& 1 + \frac{z^{2}}{2 \Tilde{L}^{2}} + \frac{5}{16}\frac{z^{4}}{\Tilde{L}^{4}} - \frac{m(1+\lambda) (d-1)}{d}\frac{z^{d}}{\Tilde{L}^{2d}}\nonumber\\
    B(z) &=& 1 - \frac{z^{2}}{2 \Tilde{L}^{2}} + \frac{5}{16}\frac{z^{4}}{\Tilde{L}^{4}} + \frac{m (1+\lambda)}{d}\frac{z^{d}}{\Tilde{L}^{2d}}~.
    \label{A(z)}
\end{eqnarray}
Similar to the planar case, it can be seen that the above form of the line element has a term that is independent of $z$ followed by $z$ dependent terms which become zero in the limit $z \to 0$. Thus, the above form of the line element is identical to the standard Fefferman-Graham form (given in eq.\eqref{FG}). 
By comparing eq.\eqref{fefferman spherical } and eq.\eqref{FG}, one can determine the coefficients of $z^0$ and $z^{d}$, and this read
\begin{eqnarray}
    g_{ab}^{(0)} dx^{a} dx^{b} = -dt^{2} + \Tilde{L}^{2} d\Omega_{d-1}^{2}~ 
\end{eqnarray}
\begin{equation}
	g_{ab}^{(d)} dx^{a} dx^{b} = \frac{m (1+ \lambda)}{d \Tilde{L}^{2d} }((d-1) dt^{2} + \Tilde{L}^{2} d\Omega_{d-1}^{2})~.
	\label{gabd spherical}
\end{equation}
The coefficient of $z^{4}$ is 
\begin{equation}
    g_{ab}^{(4)} dx^{a} dx^{b} = \frac{1}{4 \Tilde{L}^{4}}\left[\left(3 \Tilde{m} (1 + \lambda)-\frac{5}{4}\right)dt^{2} + \left(\Tilde{m}(1+\lambda) + \frac{5}{4}\right)\Tilde{L}^{2} d\Omega_{3}^{2}\right]~
    \label{bilic contradiction}
\end{equation}
where $\tilde{m}=\frac{m}{\tilde{L}^{d}}$.
It is to be observed that in the limit $\lambda \to 0$, one obtains the coefficients corresponding to a large spherical Schwarzschild black hole given in \cite{Bilic:2022psx}.\\ 
\subsection{Stress tensor for a relativistic \textbf{conformal} fluid }
In this subsection we quickly recollect the results related to the relativistic fluid stress tensor.
The long wavelength limit of AdS/CFT correspondence leads to the fluid/gravity correspondence \cite{brattan2011cft,bhattacharyya2009incompressible,minwalla2012fluid,Rangamani:2009xk,Policastro:2002se,Policastro:2002tn,Nakamura:2006ih,Baier:2007ix,Rangamani:2008gi,Janik:2006gp,Heller:2007qt,Haack:2008cp}. This duality gives us a concrete relation between gravity and physics of conformal fluids at the boundary. According to this duality, the temperature of the conformal fluid at the boundary can be calculated by assuming it as a black body. The stress tensor on the boundary is that of a viscous relativistic fluid and is given by \cite{brattan2011cft}
\begin{equation}
	T_{\mu \nu} = p g_{\mu \nu} + (\mathcal{E}+p)u_{\mu}u_{\nu}-\eta\sigma_{\mu \nu}+\cdots
\end{equation}
where $p$ is the pressure, $u_{\mu}$ is the fluid velocity, $\eta$ is the coefficient of viscosity and $\sigma_{\mu \nu}$ represents shear tensor of the fluid. The thermodynamic variables obey the following equation of state\cite{Galajinsky:2022rfu}
\begin{equation}
    p = \frac{\mathcal{E}}{(d-1)}~.
    \label{Pressure of Conformal fluid}
\end{equation}
To proceed further, we shall now assume that the conformal fluid at the AdS boundary is equivalent to the black body radiation at temperature $T$. For a black body radiation with $n_{B}$ number of massless bosons and $n_{F}$ number of massless fermions in $(d-1)$ spatial dimensions, the energy density reads 
\begin{equation}
    \mathcal{E} = \int\frac{d^{d-1}k}{(2\pi)^{d-1}}\left(\frac{n_{B} k}{e^{k/T}-1} + \frac{g n_{F} k }{e^{k/T}+1}\right) \propto T^{d}~
    \label{Energy of Conformal Fluid }
\end{equation}
where the factor $g$ is due to the degeneracy of fermions. The pressure is given by the relation in \eqref{Pressure of Conformal fluid} and reads
\begin{equation}
    p = \frac{a_{d}T^{d}}{d-1}
\end{equation}
where $a_{d}$ is a constant of proportionality. Furthermore, the entropy density of the fluid is given by
\begin{equation}
    s = \frac{(p + \mathcal{E})}{T}  = \frac{d a_{d}}{d-1}T^{d-1}~.
    \label{entropy density}
\end{equation}
\section{Thermodynamic quantities for a conformal fluid from \\gauge/gravity duality}\label{Section4}
Based upon the gauge/gravity duality, our aim now is to compute the relation between the energy density and the temperature for a conformal fluid at the boundary. We shall do this by using the definition of the holographic stress-tensor (given in eq.\eqref{gauusbonnet stress tensor}) where the gravitational input $g_{ab}^{(d)}$ is to be supplied from our obtained results given in eq.\eqref{gabd planar} and eq.\eqref{gabd spherical}.\\

\noindent For the planar black hole, we make use of eq.\eqref{gabd planar} and obtain the following
\begin{equation}
	\left<T^{a}_{~b}\right> = \left[1 + 4 (d-2) \alpha\right]\frac{(1+\lambda) r_{pl}^{d}}{ \Tilde{L}^{2d}} \mathrm{diag}(-(d-1),1,1,1,\dots)~.
\end{equation}
Reassuringly, in the limit $\alpha \to 0$, one gets back the result in \cite{Bilic:2022psx}, which reads
\begin{equation}
	\left<T^{a}_{~b}\right> =  \mathrm{diag}(-(d-1),1,1,1,\dots)\frac{r_{pl}^{d}}{L^{2d}}~.
\end{equation}
Now as we know that the $00$-component of the stress-tensor VEV gives the energy density. Hence, we have
\begin{equation}
	\mathcal{E}_{\mathrm{CF}} \equiv \left<T^{0}_{~0}\right> =\left[\frac{(d-1)}{\tilde{L}^{2d}} + \frac{(d-1)}{\tilde{L}^{2d}}(\lambda + 4(d-2)\alpha \lambda + 4(d-2)\lambda)\right]r_{pl}^d~.
	\label{approximate energy pl}
\end{equation}
From eq.\eqref{T Planar}, we note that $T_{pl}= \frac{dr_{pl}}{4\pi L^{2}}$, and by using it in the above expression, we get the relation
\begin{eqnarray}
		\mathcal{E}_{\mathrm{CF}} =\frac{(4\pi)^{d}(d-1)(1+\lambda)^{d}}{d^{d}}\left[1+\lambda \left\{1+4(d-2)(1+\alpha)\right\}\right] T_{pl}^d~.
\end{eqnarray}
Therefore, we can see that $\mathcal{E}_{\mathrm{CF}} \propto T_{pl}^d$. 
This agrees with the expected result given in eq.\eqref{Energy of Conformal Fluid } for the relativistic fluid. This is one of the main findings in this paper.
\begin{figure}[H]
	\centering
	\begin{subfigure}[t]{0.5\textwidth}
		\centering
		\includegraphics[width=8.6cm,height=6cm]{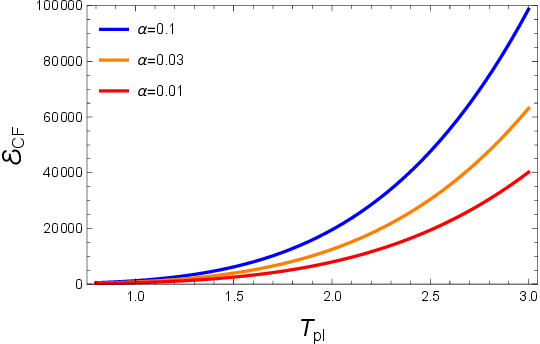}
		\caption{$d=4$}
	\end{subfigure}%
	~ 
	\begin{subfigure}[t]{0.5\textwidth}
		\centering
		\includegraphics[width=8.6cm,height=6cm]{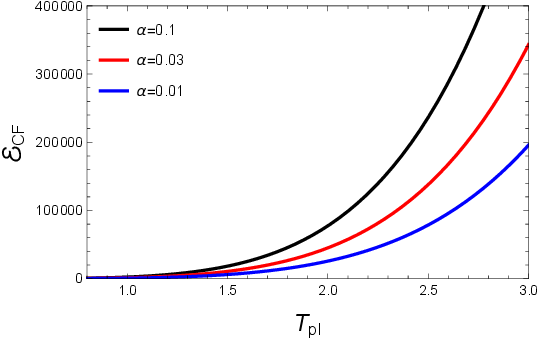}
		\caption{$d=5$}
	\end{subfigure}
	\caption{$	\mathcal{E}_{\mathrm{CF}}$ vs $T_{pl}$ plot for different values of the parameter $\alpha$.}
	\label{ecfplanar}
\end{figure}
\noindent
In the above Figure we have shown the variation of the energy density of conformal fluid with planar black hole temperature for different values of parameter $\alpha$ and $\lambda$. The $\alpha$ values are chosen from the bound in eq.\eqref{alpha bound} for a given value of $d$, and the corresponding value of $\lambda$ can be calculated from eq.\eqref{alpha reparameter}. The plot is done for $d=4$ and $d=5$ with $L=1$. From Fig.\eqref{ecfplanar} we can see that the graphs becomes more steeper for higher values of $\alpha$.\\
On the other hand, in case of the spherical black hole, we obtain
\begin{equation}
	\left<T^{a}_{~b}\right> = \frac{m(1+\lambda)}{\Tilde{L}^{2d}}\left[1 + 4 (d-2) \alpha\right]\mathrm{diag}(-(d-1),1,1,1,\dots)
\end{equation}
Similar to the planar case, in the limit $\alpha \to 0$, one gets back the result in \cite{Bilic:2022psx}, which reads
\begin{equation}
	\left<T^{a}_{~b}\right> = \frac{m}{L^{2d}}\mathrm{diag}(-(d-1),1,1,1,\dots)
\end{equation}
Once again the $00$ component of the VEV of stress tensor gives energy density of the conformal fluid, and that reads
\begin{equation}
		\mathcal{E}_{\mathrm{CF}} \equiv	\left<T^{0}_{~0}\right> = \frac{(d - 1)(1 + \lambda)}{ \Tilde{L}^{2d}}[1 + 4(d-2)\alpha]m~.
\end{equation}
Upon substituting the value of $m$ from eq.\eqref{mass parm spherical} , the above equation becomes
\begin{equation}
	\mathcal{E}_{\mathrm{CF}} \equiv	\left<T^{0}_{~0}\right> = \frac{(d - 1)(1 + \lambda)}{ \Tilde{L}^{2d}}[1 + 4(d-2)\alpha][r_{sp}^{d-4} \left(r_{sp}^{4} + L^{2}  r_{sp}^{2} + \lambda l^{4} \right)]~.
	\label{energy spherical}
\end{equation}
For a large spherical SAdS black hole $r_{sp} \gg L$, eq.\eqref{energy spherical} can be approximated as
\begin{equation}
		\mathcal{E}_{\mathrm{CF}} \approx \left[\frac{(d-1)}{\tilde{L}^{2d}} + \frac{(d-1)}{\tilde{L}^{2d}}(\lambda + 4(d-2)\alpha \lambda + 4(d-2)\lambda)\right] r_{sp}^d~.
	\label{approximate energy spherical}
\end{equation}
\begin{figure}[H]
	\centering
	\begin{subfigure}[t]{0.5\textwidth}
		\centering
		\includegraphics[width=8.6cm,height=6cm]{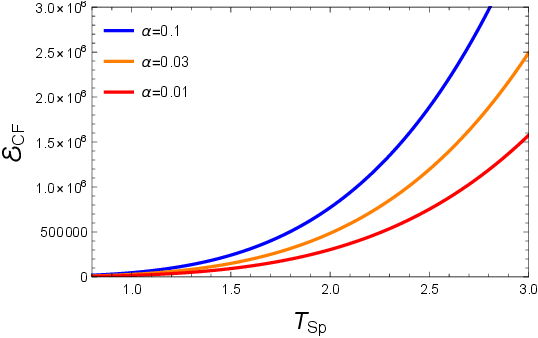}
		\caption{$d=4$}
	\end{subfigure}%
	~ 
	\begin{subfigure}[t]{0.5\textwidth}
		\centering
		\includegraphics[width=8.6cm,height=6cm]{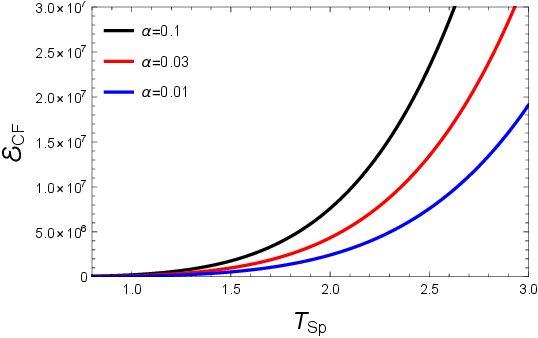}
		\caption{$d=5$}
	\end{subfigure}
	\caption{$	\mathcal{E}_{\mathrm{CF}}$ vs $T_{sp}$ plot for different values of the parameter $\alpha$.}
	\label{ecfsph}
\end{figure}
\noindent
In Figure\eqref{ecfsph}, we have plotted the variation of the energy density of the conformal fluid ($	\mathcal{E}_{\mathrm{CF}}$) with respect to the spherical black hole temperature ($T_{sp}$) for different values of the parameter $\alpha$ and $\lambda$. One should remember that $\alpha$ should be chosen from the bound in eq.\eqref{alpha bound} and the corresponding value of $\lambda$ is calculated using eq.\eqref{alpha reparameter}. The procedure of obtaining the above plot is as follows:  Using a range of values of $T_{sp}$, we have numerically solved eq.\eqref{T spherical} for $r_{sp}$, with $d=4,5$.\footnote{ It should be noted that for a given $T_{sp}$ with $d=5$, eq.\eqref{T spherical} has four roots, two of them are complex in nature and two are real. But for one of the real roots, $	\mathcal{E}_{\mathrm{CF}}$ starts decreasing for increasing value of $r_{sp}$, which is not possible as $	\mathcal{E}_{\mathrm{CF}}$ increases for increasing value of $r_{sp}$, thus this root is not physically acceptable. For $d=4$ we have three roots, one is complex and two are real in nature. Similar to the $d=5$ case we have rejected the nonphysical roots.} Then for each value of $r_{sp}$, $	\mathcal{E}_{\mathrm{CF}}$ has been calculated using eq.\eqref{approximate energy spherical}. In this way we get the allowed values of $	\mathcal{E}_{\mathrm{CF}}$ for each value of $T_{sp}$. This list has been plotted to get Fig.\eqref{ecfsph}. 

\noindent
This is another important finding in this paper. Once again, the above relation can be written down as
\begin{eqnarray}
	\mathcal{E}_{\mathrm{CF}} \approx \frac{(4\pi)^{d}(1+\lambda)^{d}L^{2d}}{d^d}\left[\frac{(d-1)}{\tilde{L}^{2d}} + \frac{(d-1)}{\tilde{L}^{2d}}(\lambda + 4(d-2)\alpha \lambda + 4(d-2)\lambda)\right] T_{sp}^d.	
\end{eqnarray}
One can also holographically compute the entropy of a conformal fluid at the boundary by using the Bekenstein-Hawking entropy of the black hole. In case of the planar black hole, this reads\cite{Reynolds:2017lwq}
\begin{equation}
	S = \lim_{\epsilon \to 0} \frac{d a_{d}}{d-1} T^{d-1} \frac{L^{d-1}}{\epsilon^{d-1}} \omega_{pl}
	\label{total BB entropy planar}
\end{equation}
where $\frac{L^{d-1}}{\epsilon^{d-1}} \omega_{pl}$ is the area of the hypersurface located at $\rho =\epsilon$, which approaches the boundary as $\epsilon\to 0$. Here $\omega_{pl}$ is $\omega_{k}$ for $k=0$.

\noindent
We now equate the above expression of entropy to the black hole entropy given in eq.\eqref{S Planar}. This in turn gives the following relation
\begin{equation}
	T^{d-1} = \frac{(d-1)}{4 d a_{d} G} \frac{\epsilon^{d-1}}{L^{d-1}} r_{pl}^{d-1}~.
\end{equation}
Now by using eq.\eqref{T Planar} in the above equation, we get the following result
\begin{align}
	T&=\left[\frac{d-1}{4d a_{d}G}\right]^{\frac{1}{(d-1)}} \frac{\epsilon}{L}\left(\frac{4\pi L^{2}}{d}\right)T_{pl}\nonumber\\
	&=4 \pi \tilde{L}^{2}(1+\lambda)\left[\frac{d-1}{4d a_{d}G}\right]^{\frac{1}{(d-1)}}\frac{\epsilon}{\tilde{L}\sqrt{1+\lambda}}T_{pl}\nonumber\\
	&=\frac{\tilde{L}}{\sqrt{1+\lambda}}\frac{4 \pi \Tilde{L}(1+\lambda)}{d} \left(\frac{d-1}{4 d a_{d} G}\right)^{\frac{1}{(d-1)}}\frac{T_{pl}}{\frac{\Tilde{L}}{\epsilon}}~.
	\label{T Cft planar}
\end{align}
\noindent
The above result is very interesting as will be clear now.
According to Tolman's relation \cite{PhysRev.35.904,PhysRev.36.1791,santiago2019tolman}, the local temperature is given by the relation
\begin{equation}
	T_{loc} = \frac{T_{BH}}{\sqrt{g_{00}}}
\end{equation}
where $T_{BH}$ is the temperature of the black hole measured by an asymptotic observer also known as the Hawking temperature. For the planar black hole temperature $T_{pl}$ in eq.\eqref{T Planar}, the local temperature measured by an observer near the $AdS$ boundary can be calculated from the above equation as follows
\begin{align}
	T_{pl}\vert^{loc} &= \frac{T_{pl}}{\sqrt{\frac{\Tilde{L}^{2}}{\rho^{2}}\left(1-(1+\lambda)\frac{\rho^{d}r_{pl}^{d}}{\Tilde{L}^{2d}}\right)}\Big|_{\rho={\epsilon}}}\\
	&= \frac{T_{pl}}{\sqrt{\frac{\Tilde{L}^{2}}{\epsilon^{2}}\left(1-(1+\lambda)\frac{\epsilon^{d}r_{pl}^{d}}{\Tilde{L}^{2d}}\right)}}~.
\end{align}
In the $\epsilon \to 0 $ limit, which is the near boundary limit, the local temperature near the AdS boundary is  $T_{pl}\vert^{loc} \approx \frac{T_{pl}}{\frac{\Tilde{L}}{\epsilon}}$.

\noindent
Using eq.\eqref{T Cft planar}, one can write down 
\begin{equation}
	T = \frac{\Tilde{L}}{\sqrt{1+\lambda}}\frac{4 \pi \Tilde{L}(1+\lambda)}{d} \left(\frac{d-1}{4 d a_{d} G}\right)^{\frac{1}{(d-1)}} T_{pl}\vert^{loc}~.
	\label{T pl loc}
\end{equation}
This is a very interesting result in this paper. We observe that the temperature of the relativistic fluid at the boundary is related to the local temperature (introduced by Tolman \cite{PhysRev.35.904,PhysRev.36.1791}) of the black hole.
On the other hand for a spherical SAdS black hole, the entropy of a conformal fluid has the following form\cite{Ramallo:2013bua}
\begin{equation}
	S = \lim_{R \to \infty} \frac{d a_{d}}{d-1} T^{d-1} R^{d-1} \omega_{sp}~
	\label{total BB entropy}
\end{equation}
where  $R^{d-1} \omega_{sp}$ is the area of the large hypersphere of radius $R$, which approaches the boundary. Here $\omega_{sp}$ is to $\omega_{k}$ for $k=1$.
We now equate the given expression of entropy to that given in eq.\eqref{Approx S Spherical }. This in turn gives the following relation 
\begin{equation}
	T^{d-1}=\frac{d-1}{4d a_{d}G}\frac{r_{sp}^{d-1}}{R^{d-1}}~.
\end{equation}
Now by using eq.\eqref{Approx T Spherical } in the above equation one can show that
\begin{equation}
	T = \frac{4 \pi \Tilde{L}(1+\lambda)}{d} \left(\frac{d-1}{4 d a_{d} G}\right)^{\frac{1}{(d-1)}}\frac{T_{sp}}{\frac{R}{\tilde{L}}} ~.
	\label{T sp loc}
\end{equation}
The local temperature of the spherical $AdS$ Gauss-Bonnet black hole is given by
\begin{align}
	T_{sp}\vert^{loc}&=\frac{T_{sp}}{\sqrt{\frac{r^2}{\tilde{L}^{2}}(1+\frac{\tilde{L}^{2}}{r^{2}}-\frac{m}{r^{d}}(1+\lambda))}\mid_{r=R}}\\
	&=\frac{T_{sp}}{\frac{R}{\tilde{L}}\sqrt{1+\frac{\tilde{L}^{2}}{R^{2}}-\frac{m}{R^{d}}(1+\lambda)}}~.
	\label{T CFT spherical}
\end{align}
In the $R\to \infty$ limit, the local temperature near the $AdS$ boundary is $T_{sp}\vert^{loc}\approx\frac{T_{sp}}{R/ \tilde{L}}$.

\noindent
Using eq.\eqref{T CFT spherical}, one can write
\begin{equation}
	T = \frac{4 \pi \Tilde{L}(1+\lambda)}{d} \left(\frac{d-1}{4 d a_{d} G}\right)^{\frac{1}{(d-1)}}T_{sp}\vert^{loc} ~.
\end{equation}
Once again we find that the temperature of the relativistic fluid is related to the Tolman's local temperature of the black hole.\\
With these results in hand, we are now in a position to the relation between the temperatures of the large spherical and planar black hole.
Comparing eq.(s) (\eqref{T sp loc},\eqref{T pl loc}), we get
\begin{equation}
    \frac{T_{sp}}{(R/\Tilde{L})} = \frac{\Tilde{L}}{\sqrt{1+\lambda}}\frac{T_{pl}}{(\Tilde{L}/\epsilon)}~.
    \label{final result 1}
\end{equation}
The above relation can be interpreted as the relation between the Hawking temperatures associated to a planar SAdS black hole and a large spherical SAdS black hole in GB gravity.
On the other hand, the local temperatures of planar and spherical SAdS Gauss-Bonnet black holes are related as
\begin{equation}
    T_{sp}\vert^{loc} \approx \Tilde{L}(1-\frac{\lambda}{2}) T_{pl}\vert^{loc}~.
    \label{final result 2}
\end{equation}
\begin{figure}[H]
	\centering
	\includegraphics[width=11cm]{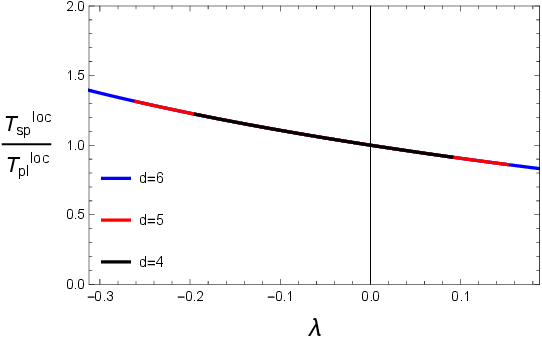}
	\caption{$T^{loc}_{sp}/T^{loc}_{pl}$ vs GB coupling coefficient ($\lambda$) plot for different values of $d$.}
	\label{ratio vs lambda}
\end{figure}
\noindent
The above equations represent the relationship between various temperatures of a planar and a large spherical Schwarzschild black hole in Gauss-Bonnet gravity. The ratio $T^{loc}_{sp}/T^{loc}_{pl}$ depends upon the GB coupling coefficient $\lambda$, with the bound of $\lambda$ depending upon the dimension. Hence the validity of eq.\eqref{final result 2} also depends upon the bound in different dimensions. In Figure \eqref{ratio vs lambda}, we have graphically shown the variation of $T^{loc}_{sp}/T^{loc}_{pl}$ with respect to $\lambda$ in various dimensions. From fig.\eqref{ratio vs lambda} it is clear that the region of validity of eq.\eqref{final result 2} expands for larger dimensions. This in turn means that one can fix the notion of temperature of a planar $AdS$ black hole in terms of the well-defined temperature of a spherical $AdS$ black hole. It is also reassuring to note that our results reduce to those in \cite{Bilic:2022psx} in the limit of the GB parameter going to zero.\\
\section{Conclusion}\label{Section5}
We now summarize our findings. In this work we have used the gauge/gravity duality to holographically calculate the vacuum expectation value (VEV) of the stress tensor associated to a conformal fluid (at the boundary) by considering the bulk theory to be the Gauss-Bonnet (GB) gravitational theory. Using the fact that the holographic stress tensor is related to the asymptotic form of the bulk metric, we first transform the line element to the Fefferman-Graham form. By incorporating this we then compute the holographic stress tensor, energy density and entropy density for both large spherical and planar GB AdS Schwarzschild (SAdS) black hole solutions. Our results are in conformity with those in the literature that finite coupling corrections appear in the expressions of energy density and entropy density in case of GB gravity. We then compare the obtained results of energy density with that of a conformal fluid at the boundary. From the correspondence between the entropy of the CFT and black hole entropy, we then proceed to express the temperature of the conformal fluid in terms of the Hawking temperature of the black hole, which enables us to establish the relationship between $T$, $T_{sp}$ and $T_{pl}$ which corresponds to the temepratures of the conformal fluid, large spherical GB SAdS black hole and planar GB SAdS black hole respectively.
We have noticed that the Hawking temperature of a large spherical GB SAdS black hole and planar GB SAdS black hole is very high (as horizon radius is much greater than the AdS radius), but due to a very large redshift near the AdS boundary the actual horizon temperature is unobservable. So for an observer near the AdS boundary, the GB SAdS black hole appears to be lukewarm. A similar phenomenon was shown for the case of a usual AdS Schwarzschild black hole in \cite{hubeny2010hawking}, although the expression for the redshift factors of the GB black hole and the Schwarzschild black hole are different. In the limit of the GB coupling parameter tending to zero, the redshift factor in our analysis exactly matches with the one obtained for the usual Schwarzschild case. Finally, comparing the entropy of the planar and large spherical black holes with the entropy of the conformal fluid we relate the Hawking temperatures of the mentioned black hole solutions. Furthermore, by incorporating the Tolman correction, we also relate the local temperatures (Tolman temperatures) of both the black holes. The effects of the GB parameter ($\lambda$) in this set up is easy to note from the relations given in eq.(s)(\eqref{final result 1},\eqref{final result 2}). 
The relation between the Tolman temperatures of a planar and a large spherical GB SAdS black holes in turn helps us to remove the horizon temperature ambiguity for a SAdS black hole in GB gravity. 
This study has gone some way towards enhancing our understanding about the finite coupling corrections on the temperature of a conformal fluid and its relation with both Hawking and Tolman temepratures of a black hole in Gauss-Bonnet (AdS) gravity. On the other hand, this study also act as a stepping stone to remove the horizon temperature ambiguity of a planar GB SAdS black hole.  
\section{Acknowledgments}
SP would like to thank SNBNCBS for the Junior Research Fellowship. AS would like to thank S.N. Bose National Centre for Basic Sciences for the financial support through its Advanced Postdoctoral Research Programme.

\bibliographystyle{hephys.bst}
\bibliography{reference.bib}
\end{document}